\documentclass[aps,prb, preprint, superscriptaddress, amsmath,amssymp,showpacs]{revtex4-1}
\usepackage{graphicx}
\usepackage{amsfonts}
\usepackage{amsmath}
\usepackage[usenames,dvipsnames]{color}
\usepackage{bm}
\usepackage{amssymb}

\usepackage{multirow}
\usepackage{epstopdf}
\usepackage{natbib}
\usepackage{ulem}

\def\jcp#1#2#3{J.~Chem.~Phys.~{\bf #1},\ #2\ (#3)}
\def\cpl#1#2#3{Chem.~Phys.~Lett.~{\bf #1},\ #2\ (#3)}

\def\pra#1#2#3{Phys.~Rev.~A~{\bf #1},\ #2\ (#3)}
\def\prl#1#2#3{Phys.~Rev.~Lett.~{\bf #1},\ #2\ (#3)}

\def\jpb#1#2#3{J. Phys. B: At. Mol. Opt. Phys. {\bf #1},\ #2\ (#3)}

\def\k1{k_1}
\def\k2{k_2}
\def\q1{q_1}
\def\q2{q_2}

\def\({\left (}
\def\){\right )}
\def\[{\left [}
\def\]{\right ]}

\newcommand{\beq}{\begin{equation}}
\newcommand{\eeq}{\end{equation}}

\newcommand{\threejm}[6]{ \left(\begin{array}{ccc} #1 & #3 & #5\\
                                              #2 & #4 & #6
                                \end{array}
                          \right)}

\begin{document}
\date{\today}
\flushbottom \draft
\title{Restricted basis set coupled-channel  calculations on atom-molecule collisions in magnetic fields} 

\author{Masato Morita}
\author{Timur V. Tscherbul}
\affiliation{Department of Physics, University of Nevada, Reno, NV, 89557, USA} \email[]{ttscherbul@unr.edu}

\begin{abstract}
Rigorous coupled-channel quantum scattering calculations on molecular collisions in external fields are computationally demanding due to the need to account for a large number of coupled channels and multiple total angular momenta $J$ of the collision complex. We show that by restricting the number of total angular momentum basis states to include only the states with helicities $K\le K_\text{max}$  it is possible to obtain accurate elastic and inelastic cross sections for He~+~CaH, Li~+~CaH and Li~+~SrOH collisions at a small fraction of the computational cost of the full coupled-channel calculations (where $K$ is the projection of the molecular rotational angular momentum on the atom-diatom axis).  
The optimal size of the  truncated helicity basis set depends on the mechanism of the inelastic process and on the magnitude of the external magnetic field. For dipolar-mediated spin relaxation in ultracold Li~+~CaH and Li~+~SrOH collisions, we find that a minimal helicity basis set ($K_\text{max}=0$) gives quantitatively accurate results at ultralow collision energies,
leading to nearly 90-fold gain in computational efficiency.
Larger basis sets are required to accurately describe the resonance structure in Li~+~CaH and Li~+~SrOH inelastic cross sections in the few partial wave-regime ($K_\text{max}=3$) as well as  indirect spin relaxation in He~+~CaH collisions ($K_\text{max}=1$). Our calculations indicate that the resonance structure is due to an interplay of the spin-rotation and Coriolis couplings between the basis states of different $K$ and the couplings between the  rotational states of the same $K$ induced by the anisotropy of the interaction potential.

\end{abstract}

\maketitle
\clearpage
\newpage

\section{Introduction}

Steady experimental advances in molecular cooling and trapping \cite{ColdMoleculeBook09,ColdMoleculeBook17,Science17,njp09,KRb08} have led to the production of trapped  ensembles of molecules in precisely defined internal states at temperatures below 1~K,  which are being actively used  to study molecular collision dynamics at ultralow temperatures \cite{KRb10,JunARPC14,RempeScience}. These studies explore the new regimes  of quantum dynamics intrinsic to low-temperature molecular scattering, including Wigner threshold laws, scattering resonances \cite{AlexanderUncloaking,ChefdevilleO2H2,Narevicius16}, and tunnelling \cite{BalaAlexCPL01,BalaJCPreview16,prl15}. Another remarkable aspect of ultracold molecular collisions  is the capability to control their outcome with external electromagnetic fields \cite{KremsDalgarno04,VolpiBohn02,prl06,GonzalezMartinezHutson07,njp09,RomanColdControlledChemistry08,ColdMoleculeChapter17,prl15,BalaJCPreview16}.

A rigorous theoretical description of two-body collisions in ultracold molecular gases in the presence of external fields is essential for interpreting the results of  experimental measurements on trapped molecules and atom-molecule mixtures \cite{njp09,ColdMoleculeChapter17}. While elastic collisions between trapped molecules are desirable as they lead to thermalization, inelastic collisions  release the energy stored in the molecule's internal degrees of freedom, thereby reducing the efficiency of sympathetic and evaporative cooling of molecular gases \cite{njp09,ColdMoleculeChapter17}. It is therefore essential to be able to estimate the low-temperature collisional properties of molecular gases and atom-molecule mixtures from first principles based on the solution of  multichannel Schr\"odinger equation for the collision complex \cite{njp09,ColdMoleculeChapter17,prl15,BalaJCPreview16}.

The numerical solution of the multichannel Schr{\"o}dinger equation  for molecular collisions in the presence of external fields poses a challenging computational problem due to an extremely large density of states inherent to most molecular collision complexes \cite{RbOH06,HutsonSympatheticCooling15,pra11,pra17,ColdMoleculeChapter17}. This problem arises  due to the strong and anisotropic intermolecular interactions, which couple a large number of low-energy rotational channels \cite{RbOH06,pra11,HutsonSympatheticCooling15,ColdMoleculeChapter17,pra17}. Taking advantage of the symmetry of the collision complex by using the eigenstates of its  total angular momentum as a basis provides an efficient way to handle the anisotropic interactions, leading to a dramatic reduction in the number of scattering channels \cite{ColdMoleculeChapter17,jcp10}. The  ensuing reduction of the computational cost has opened up the possibility to perform converged coupled-channel (CC) calculations on heavy, strongly anisotropic atom-molecule collisions and chemical reactions in the presence of external magnetic \cite{pra11,pra17,pra18subm} and electric \cite{pra12,prl15} fields. However, such calculations  still involve thousands of scattering channels, calling for the development of new methods and approximations to further reduce the number of scattering basis functions.

The use of incomplete helicity basis sets is one such approximation, which takes advantage of an approximate conservation of molecular helicity $K$, the projection of the molecule's rotational angular momentum on the atom-molecule axis \cite{Launay90,BrunoLepetit91,RomanPRCC,BalaNJP15}. This near-conservation ensures that only a limited number of $K$ values make a non-negligible contribution to the scattering process, and hence the basis set can be truncated at $K=K_\text{max}$ without compromising the quality of the results. Restricted helicity basis sets are commonly used in field-free quantum CC calculations of inelastic collisions \cite{RomanPRCC} and chemical reactions \cite{Launay90,BrunoLepetit91,ABC} of closed-shell diatomic molecules with structureless atoms.
We note that for chemical reactions that are not dominated by near-collinear transition states such as Li~+~HF $\to$ LiF~+~H, such basis sets  provide no computational advantage, and complete helicity basis sets should be used to obtain quantitatively accurate results \cite{BalaNJP15}. 

A related approximate technique, which retains all values of $K$ in the basis set but 
neglects the Coriolis coupling between them, is known as the helicity-conserving coupled-states (CS) approximation \cite{Kouri,Pack}. The CS approximation generally provides quantitatively accurate results for molecular collisions at room temperature and above \cite{Kouri,Pack,Roman01,JacekLiqueChapter}. The CS approximation has also been tested for cold molecular collisions and reactions in the absence of external fields and found to produce accurate results  \cite{JacekLiqueChapter, BalaFHCl} for the background (non-resonant) reactive scattering of HCl and DCl  molecules  in their ground rotational states with F atoms. However, neither helicity truncation nor CS approximation have been applied to molecular collisions in the presence of external electromagnetic fields.

Here, we explore the use of restricted helicity basis sets in quantum scattering calculations on ultracold atom-molecule collisions in the presence of an external magnetic field. We focus on an important class of inelastic collisions that change the internal spin (Zeeman) state of the colliding molecules known as spin relaxation. We show that quantum scattering calculations employing reduced helicity basis sets provide accurate results for the elastic and spin relaxation cross sections in the resonance-free regime over a wide range of magnetic fields  and collision energies at a reduced computational cost.

This paper is organized as follows. In Sec.~II we outline the theoretical methodology for rigorous atom-molecule CC calculations in a magnetic field and describe the procedure of basis set truncation. In Sec.~III we compare the approximate results obtained using truncated basis sets with the full CC calculations for low-temperature He~+~CaH, Li~+~CaH, and Li~+~SrOH collisions in a magnetic field. We analyze the performance of the different $K$ basis sets and  estimate the computational speedups achieved. In Sec.~IIIB, we examine the resonance structure in Li~+~CaH inelastic cross sections at high magnetic fields as an example of the situation where truncated basis calculations do not provide an accurate approximation to the full CC results. Sec.~IV concludes with a brief summary of results and outlines directions for future research.

\section{Theory}

In this section we first outline the rigorous CC  methodology for atom-molecule collisions in the presence of an external magnetic field \cite{jcp10} and then describe the procedure of helicity basis set truncation.
Throughout this work we will be concerned with non-reactive collisions of diatomic molecules in the electronic states of $^2\Sigma$ symmetry with atoms in the $^1\text{S}$ and  $^2\text{S}$ electronic states. Examples include  collisions of He($^1\text{S}$) atoms with CaH($^2\Sigma$) molecules,  Li($^2\text{S}$) atoms with CaH and SrOH($^2\Sigma$) molecules, and Rb($^2\text{S}$) atoms with SrF($^2\Sigma$) molecules. Previous theoretical studies have found low inelastic collision rates  in these systems, suggesting favorable prospects for sympathetic cooling of CaH, SrOH, and SrF molecular radicals by ultracold alkali-metal atoms in a magnetic trap \cite{pra11,pra17,pra18subm}.

The Hamiltonian of the atom-molecule collision complex is conveniently expressed in a body-fixed (BF) coordinate frame with the $z$-axis defined by the atom-molecule Jacobi vector $\bm{R}$ and the $y$-axis perpendicular to the collision plane \cite{jcp10,vdwReview,ParkerPack87}
\begin{equation}\label{Hbf}
\hat{H} = -\frac{1}{2\mu R}\frac{\partial^2}{\partial R^2} R + \frac{1}{2\mu R^2}  (\hat{\bm{J}} - \hat{\bm{N}} - \hat{\bm{S}}_a - \hat{\bm{S}})^2 + \hat{V}(R,\theta) +  \hat{H}_\text{at} + \hat{H}_\text{mol},
\end{equation}
where $\bm{r}$ is the internuclear Jacobi vector of the diatomic molecule, 
$R=|\bm{R}|$, $r=|\bm{r}|$, and $\theta$ is the angle between $\bm{R}$ and $\bm{r}$. In Eq. (\ref{Hbf}), $\hat{\bm{S}}_a$ and $\hat{\bm{S}}$ stand for the electron spins of the atom and the molecule, $\mu$ and $\hat{\bm{J}}$  are the reduced mass and the total angular momentum of the atom-molecule collision complex, $\bm{N}$ is the rotational angular momentum of the diatomic molecule (see below) and $\hat{V}(\bm{R},\bm{r})$ is the atom-molecule interaction potential, including both the electrostatic and magnetic dipole-dipole interactions (for $S_a\ne 0$).

The asymptotic Hamiltonian $\hat{H}_\text{as}=\hat{H}_\text{at} + \hat{H}_\text{mol}$ describes the threshold structure of the collision complex in the limit $R\to\infty$. The molecular Hamiltonian  $\hat{H}_\text{mol}$ defines the energy levels of the $^2\Sigma$ diatomic molecule in its electronic and vibrational ground states in the presence of an external magnetic field $B$ \cite{Mizushima,BrownCarrington}
\begin{equation}\label{Has2Sigma}
\hat{H}_\text{mol} = B_e \hat{\bm{N}}^2 + \gamma\hat{\bm{N}}\cdot\hat{\bm{S}} + 2\mu_0 {B} \hat{S}_{Z}
\end{equation}
where $B_e$ is the rotational constant of the molecule, $\hat{\bm{N}}$ is the rotational angular momentum, $\hat{\bm{S}}$ is the electron spin, $\hat{S}_{Z}$ is the projection of  $\hat{\bm{S}}$ on the magnetic field axis (the space-fixed (SF) quantization axis), $\mu_0$ is the Bohr magneton, and $\gamma$ is the   spin-rotation constant \cite{BrownCarrington}. In what follows we will assume that the internuclear distance in the diatomic molecule $r$ is fixed at its equilibrium value $r_e$, which is a good approximation  if the atom-molecule interaction potential $\hat{V}(R,\theta)$ depends on $r$ only weakly (true for He-CaH \cite{HeCaHpes} and alkali-SrF interactions \cite{PiotrMaciej17}). Finally, the Hamiltonian of the $^2$S atom is given by (neglecting the hyperfine structure)
\begin{equation}\label{Hatom}
\hat{H}_\text{at} = 2\mu_0 {B} \hat{S}_{a_Z}
\end{equation} 
where the operator $\hat{S}_{a_Z}$ gives the projection of  $\hat{S}_a$ on the SF quantization  axis.
Equation (\ref{Has2Sigma}) represents the simplest possible Hamiltonian for an open-shell molecular radical; however, additional spin-dependent terms may be added to it as necessary to account for the  electronic states of $^3\Sigma$ or $^2\Pi$ symmetries \cite{HeOH, jcp10} or the hyperfine structure \cite{pra_ybf07,MaykelHyperfine11,jcp10}.


To solve the time-independent  Schr\"odinger equation with the Hamiltonian (\ref{Hbf}) we expand the wavefunction of the atom-molecule collision complex in BF basis functions \cite{jcp10,vdwReview,ParkerPack87}
\begin{equation} \label{BFexpansion}
|\Psi\rangle = \frac{1}{R}\sum_{\alpha,J,\Omega} F^M_{\alpha J\Omega}(R) |\alpha\rangle |JM\Omega\rangle, 
\end{equation}
\begin{equation}\label{BFbasis}
 |\alpha\rangle  = |NK\rangle |S\Sigma\rangle |S_a\Sigma_a\rangle,
\end{equation}
where $\Omega$, $K$,  $\Sigma$, and $\Sigma_a$ are the projections of $\hat{J}$, $\hat{N}$,  $\hat{S}$, and  $\hat{S}_a$ on the BF quantization axis $z$ (denoted collectively  by $\alpha$); $\Omega=K+\Sigma+\Sigma_a$. The Wigner $D$-functions $|JM\Omega\rangle=\sqrt{(2J+1)/8\pi^2}D^{J*}_{MK}(\hat{\Omega})$ depend on the Euler angles which specify the position of BF axes $x$, $y$, and $z$ in the SF frame.  The functions $|NK\rangle=\sqrt{2\pi}Y_{NK}(\theta,0)$,  $|S\Sigma\rangle$, and $|S_a\Sigma_a\rangle$ describe the orientation of the diatomic molecule and the spin degrees of freedom in the BF frame.

The radial expansion coefficients $F^M_{\alpha J\Omega}(R)$, which carry information about scattering observables, may be found by solving the CC equations  \cite{jcp10} 
\begin{align}\label{CC}\notag
\left[ \frac{d^2}{dR^2} +2\mu E\right] F^M_{\alpha J\Omega}(R)=2\mu \sum_{\alpha',J,'\Omega'} \langle \alpha J\Omega |  \hat{V}(R,\theta) &+ \frac{1}{2\mu R^2} (\hat{\bm{J}} - \hat{\bm{N}} - \hat{\bm{S}}
-\hat{\bm{S}}_a  )^2 \\&+\hat{H}_\text{mol} + \hat{H}_\text{at} | \alpha'J'\Omega'\rangle F^M_{\alpha' J'\Omega'}(R),
\end{align}
where $E$ is the total energy. 
The  matrix elements of the interaction potential, orbital angular momentum and the asymptotic Hamiltonian on the right-hand side of Eq.~(\ref{CC}) can be evaluated as described in our previous work \cite{jcp10,pra18subm}. We use the accurate spectroscopic constants of CaH($^2\Sigma^+$) and SrOH($^2\Sigma^+$) and the most recent {\it ab initio}  interaction potentials for He-CaH \cite{HeCaHpes}, Li-CaH \cite{pra11}, and Li-SrOH \cite{pra17}. The size of the basis set is controlled by the truncation parameters $J_\text{max}$ and $N_\text{max}$, which give the maximum  values of the total angular momentum $J$ and rotational angular momentum $N$ in the basis set.

We integrate the CC equations using the modified log-derivative method \cite{DavidManolopoulos86} to obtain the log-derivative of the multichannel scattering wavefunction in the asymptotic  region, where the atom-molecule interaction potential is negligible compared to the centrifugal potential and to the collision energy. 
The log-derivative matrix is then transformed to the SF representation and to a basis, which diagonalizes the asymptotic Hamiltonian \cite{jcp10}. As a final step, matching the transformed log-derivative matrix to the Riccati-Bessel functions and their derivatives \cite{Johnson73,jcp10} yields  the $S$-matrix elements, from which the scattering cross sections are obtained using standard expressions \cite{jcp10}. 
The calculated scattering cross sections are converged to $<10\%$ with respect to the basis set and radial grid parameters. Table~I lists the values of the convergence parameters for the atom-molecule collision systems studied in this work.


To describe truncated helicity basis sets, we introduce an additional parameter  $K_\text{max}$, which gives the maximum value of the helicity in the BF basis (\ref{BFbasis}). 
The  maximum possible  value of $K_\text{max}$ is given by $\min(N_\text{max},J_\text{max}$
$+S+S_a$), which corresponds to full CC calculations.
Table~II lists the numbers of BF basis functions for the atom-molecule collision systems studied here as a function of $K_\text{max}$. The size of the basis set decreases substantially with decreasing $K_\text{max}$, leading to a  substantial reduction of the number of scattering channels at low $K_\text{max}$. Since the computational cost of CC calculations scales as $N^3$  with the number of scattering channels $N$, using basis sets with small $K_\text{max}$ can result in nearly two orders of magnitude reduction of the computational cost. 

A priori, it is not clear whether truncation of the full helicity basis set can lead to quantitatively accurate results for low-temperature atom-molecule collisions in a magnetic field. Indeed,  limiting the number of $K$-states in the basis set affects the matrix elements of $K$-dependent interactions in the molecular Hamiltonian, such as
 the spin-rotation interaction \cite{jcp10}
\begin{multline}\label{SRint}
 \langle NK | \langle S \Sigma|  \gamma\hat{\bm{N}}\cdot\hat{\bm{S}} |N'K'\rangle |S\Sigma'\rangle = \delta_{NN'} \bigl{[} \gamma_\text{SR} K\Sigma\delta_{KK'}\delta_{\Sigma\Sigma'}  \\ + \frac{\gamma_\text{SR}}{2} [N'(N'+1) - K'(K'\pm 1)]^{1/2} [S(S+1) - \Sigma'(\Sigma'\mp 1)]^{1/2} \delta_{K,K'\pm 1}\delta_{\Sigma,\Sigma'\mp1} \bigr{]},
\end{multline}
the atom-molecule interaction potential
\begin{multline}\label{meV}
\langle JM\Omega | \langle NK | \langle S \Sigma| V(R,\theta) | J'M'\Omega'\rangle |N'K'\rangle |S\Sigma'\rangle  = \delta_{JJ'}\delta_{MM'}\delta_{\Sigma\Sigma'} \\ \times [(2N+1)(2N'+1)]^{1/2} (-1)^{K} \sum_{\lambda} V_\lambda(R) \threejm{N}{-K}{\lambda}{0}{N'}{K'} \threejm{N}{0}{\lambda}{0}{N'}{0},
\end{multline} 
and the squared orbital angular momentum of the atom-molecule complex
  \begin{multline}\label{meL2}
  \langle JM\Omega | \langle NK | \langle S \Sigma| (\hat{\bm{J}}-\hat{\bm{N}}-\hat{\bm{S}})^2 | J'M'\Omega'\rangle |N'K'\rangle |S\Sigma'\rangle 
 =  \delta_{JJ'}\delta_{MM'}\delta_{NN'} \\  \times [
  \mathcal{C}_1(J,\Omega,\alpha) \delta_{\Omega\Omega'}\delta_{KK'}\delta_{\Sigma\Sigma'} 
 -  \mathcal{C}_2 (J,\Omega,\Omega',\alpha,\alpha') \delta_{\Omega\Omega'} \delta_{\Omega,\Omega'\pm1} \delta_{K,K'\pm1} \delta_{\Sigma\Sigma'} \\
  - \mathcal{C}_3 (J,\Omega,\Omega',\alpha,\alpha')  \delta_{\Omega,\Omega'\pm1} \delta_{K,K'} \delta_{\Sigma,\Sigma'\pm1} 
  - \mathcal{C}_4 (\alpha,\alpha')   \delta_{\Omega\Omega'} \delta_{K,K'\pm1} \delta_{\Sigma,\Sigma'\mp1}
\biggr{]}
\end{multline}
where  $\mathcal{C}_i$ are angular momentum coefficients that generally depend on $J$, $\Omega$, $N$, $\Sigma$, and $K$  (see Ref.~\citenum{jcp10} for explicit expressions).
Note that {the spin-rotation interaction (Eq.~\ref{SRint}) as well as the centrifugal term (Eq.~\ref{meL2}) couple the basis functions with $K'=K$ and $K'=K\pm 1$ whereas the atom-molecule interaction potential (Eq.~\ref{meV}) is diagonal in $K$. \cite{jcp10} } 

The above expressions suggest  that truncation of the helicity basis is expected to produce quantitatively accurate results if the 
 matrix elements of the spin-rotation interaction, the atom-molecule interaction potential, and of the centrifugal term omitted from the basis are much smaller in absolute magnitude than the matrix elements retained in the basis.
In the following we will examine the effect of  basis set truncation on the scattering observables for ultracold He~+~CaH, Li~+~CaH, and Li~+~SrOH collisions in a magnetic field.

\section{Results and Discussion}

\subsection{He~+~CaH collisions}

In this section, we present the results of restricted basis set CC calculations of the elastic and spin relaxation  cross sections for He~+~CaH collisions in the presence of an external magnetic field. He~+~CaH is a prototype weakly anisotropic collision system explored in a number of previous theoretical studies \cite{Roman03,Roman03a,prl06,jcp10} using rigorous CC methods and an accurate {\it ab initio} PES. These theoretical calculations were in semi-quantitative  agreement with experimental measurements of collision-induced spin relaxation rates of magnetically trapped CaH molecules at 0.5~K \cite{CaHtrappingNature98}. 

\begin{figure}[h]
	\centering
	\includegraphics[width=1.\textwidth, trim = 0 0 0 0]{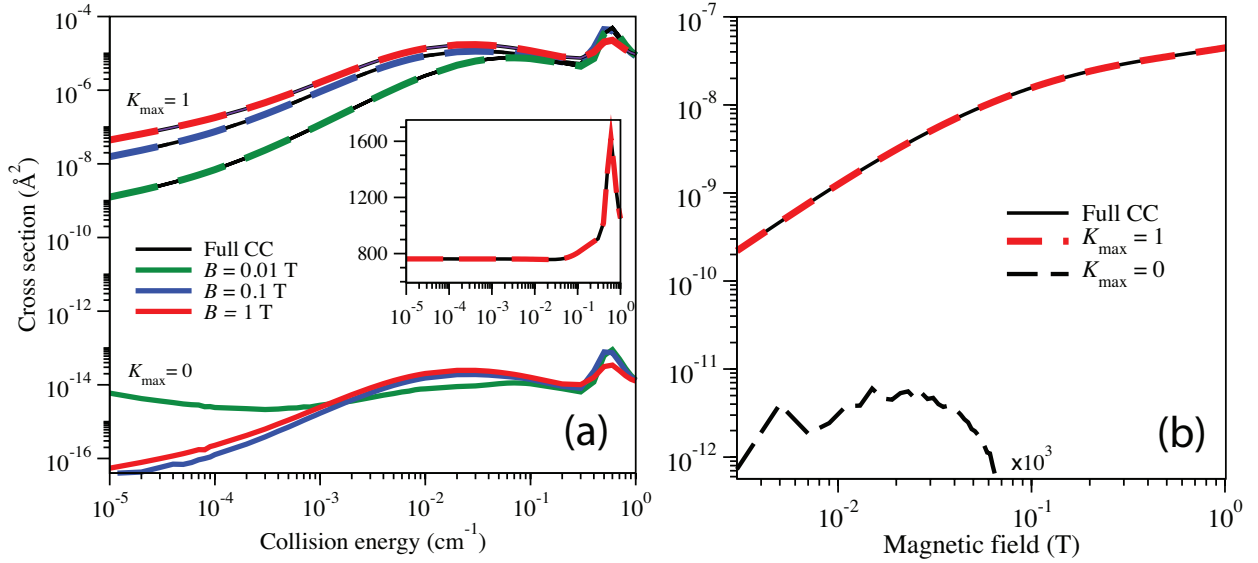}
	\renewcommand{\figurename}{Fig.}
	\caption{(a) Spin relaxation cross sections for He~+~CaH collisions as a function of the collision energy.  Top lines: full CC results (solid lines) compared with restricted basis set calculations (dashed lines) with $K_\text{max}=1$ for $B=1$~T (top), $B=0.1$~T (middle), $B=0.01$~T (bottom). Bottom lines: restricted bases set results for $K_\text{max}=0$.  The inset shows the elastic cross section as a function of the collision energy at $B=0.01$ T.  (b) Magnetic field dependence of spin relaxation cross sections for He~+~CaH  at  $E_C=10^{-5}$ cm$^{-1}$. Full CC results (solid line) are compared with restricted basis set results (dashed lines) calculated with $K_\text{max}=1$ (red/grey line) and $K_\text{max}=0$ (black line).}
\label{fig:hecahE}	
\end{figure}

Figure~\ref{fig:hecahE}{(a)} shows the cross sections for elastic scattering and inelastic relaxation in He+CaH  collisions as a function of the collision energy. We focus on the  spin-flipping  transition 
$M_{S}=1/2\to M_{S}=-1/2$ in collisions of CaH($^2\Sigma$) molecules within the ground rotational state, which was the subject of several experimental  \cite{MolPhys13,CaHtrappingNature98} and many theoretical \cite{Roman03,Roman03a,prl06,jcp10} studies. Here,  
{$M_{S}$}
is the projection of the molecular spin onto the magnetic field axis.

The inelastic cross sections are 6 orders of magnitude  smaller than the elastic cross section shown in the inset of Fig.~\ref{fig:hecahE} and they decrease with decreasing collision energy before reaching the Wigner threshold regime \cite{jcp10,Roman03}  $\sigma_\text{inel}\propto E^{-1/2}$ . We observe that  the inelastic cross sections computed with $K_\text{max}=1$ are in excellent agreement with exact CC results to within 1\%. In  contrast, restricted basis set results obtained with $K_\text{max}=0$ underestimate the exact cross sections by more than 5 orders of magnitude. As shown below, this striking difference is due to a peculiar mechanism of spin relaxation  in He+CaH collisions.  The elastic cross sections  agree well with full CC calculations for both $K_\text{max}=0$ and 1.

Figure~\ref{fig:hecahE}(b) shows the magnetic field dependence of the cross sections for spin relaxation in  He+CaH collisions in the $s$-wave scattering regime ($E_C=10^{-5}$~cm$^{-1}$). The CC cross sections grow monotonously with the Zeeman splitting between the $M_S=\pm 1/2$ levels of CaH due to an enhanced rate of tunnelling through the centrifugal barrier in the outgoing collision channel \cite{VolpiBohn02}. The restricted basis set calculations with $K_\text{max}=1$ are in excellent agreement with the accurate CC results regardless the entire range of magnetic field explored  here (0.003~T - 1~T). However, the  minimal basis set calculations with $K_\text{max}=0$ underestimate the inelastic cross sections by 5-7 orders of magnitude 

The inability of the minimal basis set ($K_\text{max}=0$) to provide reliable results for the spin relaxation transition in CaH indicates that $K=1$ basis functions play an essential role in the spin relaxation mechanism. Indeed, as shown by Krems and co-workers \cite{Roman03a}, spin relaxation in collisions of $^2\Sigma$ molecules with structureless atoms is a third-order  process mediated by the anisotropy of the atom-molecule interaction potential and by the spin-rotation interaction within the excited rotational state manifold. The spin-rotation interaction flips the molecular spin projection $\Sigma$ while changing the value of $K$ by one so as to conserve the BF projection of the total angular momentum {$\Omega=$} 
$K+\Sigma$.
A proper description of  this process thus requires  basis  functions with $K= 1$, which are missing from the minimal  $K_\text{max}=0$ basis set, leading to inaccurate results.

\subsection{Li~+~CaH collisions}

Unlike the He~+~CaH just considered,  Li~+~CaH is a prototype of a strongly bound,  anisotropic collision complex with spin-dependent interactions. Ultracold Li$(\uparrow)$~+~CaH($\uparrow$) collisions in a magnetic field are predominantly elastic \cite{pra11} despite the presence of strongly anisotropic interactions, which indicate good prospects for sympathetic cooling of magnetically trapped 
CaH$(N=0)$ molecules by ultracold Li atoms \cite{pra11}.  In the following, we will use arrows to denote the atomic and molecular Zeeman states  $\uparrow$ ($\downarrow$)  corresponding to $M_{S_i}=+1/2$ ($-1/2$). For instance, collisions of CaH molecules with Li atoms initially in their fully spin-polarized initial states $M_{S_a}=M_{S}=1/2$  are denoted as Li($\uparrow$)~+~CaH($\uparrow$).

\begin{figure}[h]
	\centering
	\includegraphics[width=1.1\textwidth, trim = 100 90 0 0]{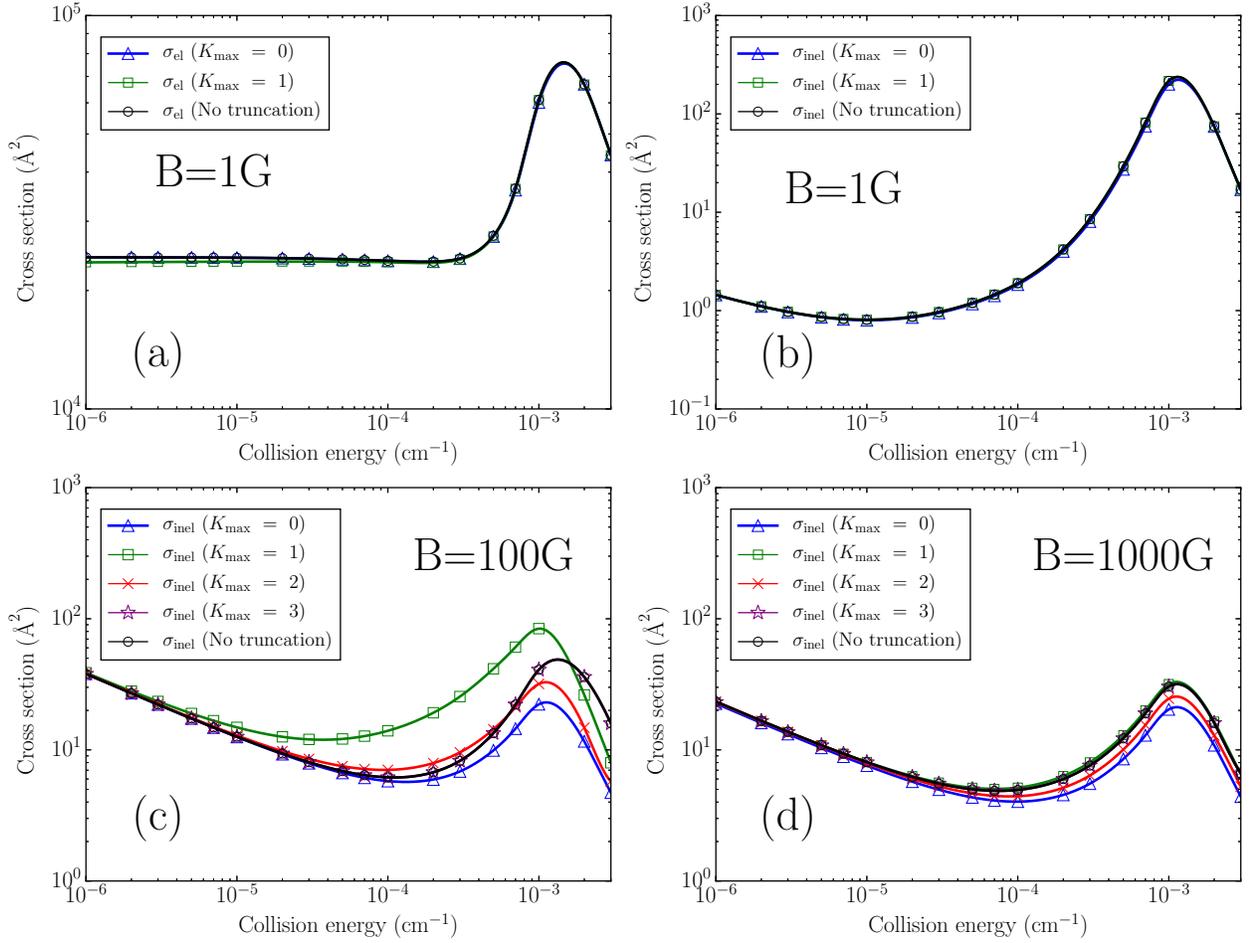}
	\renewcommand{\figurename}{Fig.}
	\caption{Elastic (a) and inelastic spin relaxation (b)-(d) cross sections for Li$(\uparrow)$~+~CaH$(\uparrow)$ collisions as a function of the collision energy for $B=1$~G (a)-(b), $B=100$~G (c), and $B=1000$~G (d). Full CC results (circles) are compared with restricted basis set calculations with $K_\text{max}=0$ (triangles), 1 (squares), 2 (crosses), and 3 (stars).}
	\label{fig:licahE}
\end{figure}

Figure~\ref{fig:licahE}(a) shows the collision energy dependence of the elastic  cross section, which approaches a constant value in the $s$-wave regime ($E_C<10^{-4}$ cm$^{-1}$) and displays a shape resonance at $E_C\sim 1.5\times 10^{-3}$ cm$^{-1}$ (2 mK). 
The cross section computed using the maximally truncated basis 
($K_\text{max}= 0$) agrees well with full CC results over a wide  range of collision energies spanning four orders of magnitude. The agreement is particularly good in the multiple partial-wave regime near the shape resonance, which suggests that single-channel shape resonances can be described with quantitative accuracy by CC calculations including only the lowest value of $K=0$ in the basis set. As  Table~II shows, such calculations are nearly {90}  times more computationally efficient than the full CC calculations.


Figures~\ref{fig:licahE}(b)-(d) show the collision energy dependence of the inelastic cross sections calculated at several representative magnetic fields. While at the lowest field of 1~G the cross sections calculated with the minimal basis set ($K_\text{max} = 0$) are indistinguishable from exact CC results, significant deviations are observed in the intermediate-field regime ($B=100$ G). In particular, $K_\text{max}=1$ calculations strongly overestimate the full CC results in the vicinity of the 2 mK shape resonance. As expected, the  results computed with restricted basis sets gradually converge to the full CC limit with increasing $K_\text{max}$, although as illustrated in Fig.~\ref{fig:licahE}(b) the convergence can be slow. We attribute this to the presence of a scattering resonance in the inelastic cross section, which occurs in the multiple partial-wave regime near $B=100$~G. 
As shown below, the resonance is mediated  by the 
{spin-rotation}
and the centrifugal couplings, which connect basis functions with  $K'=K\pm 1$,
and the matrix elements of the interaction potential, which is diagonal in $K$.

As shown in our previous work \cite{pra11}, spin relaxation in doubly polarized Li($\uparrow$)+CaH($\uparrow$) collisions occurs through an interplay of two mechanisms. The first indirect mechanism discussed in Sec.~IIB above involves the spin-rotation interaction in the excited rotational states of CaH and the anisotropy of the Li-CaH interaction potential. The second mechanism is a direct relaxation process mediated by the magnetic dipole-dipole interaction \cite{pra17,pra18subm} between 
{Li($\uparrow$) and CaH($\uparrow$).}
Because the magnetic dipole-dipole interaction is diagonal in $K$ (see, {\it e.g.}, Eq. (8) of Ref.~\citenum{pra18subm}) the direct mechanism does not require a change in the value of $K$, in contrast to  the indirect mechanism discussed above, which does. The rapid convergence of restricted basis set calculations to the exact result in the ultracold regime shown in Fig.~\ref{fig:licahE}(b)-(d) is then due to the dominance of the direct spin relaxation mechanism in this regime \cite{pra11}.

\begin{figure}[h]
	\centering
	\includegraphics[width=1.1\textwidth, trim = 100 80 0 0]{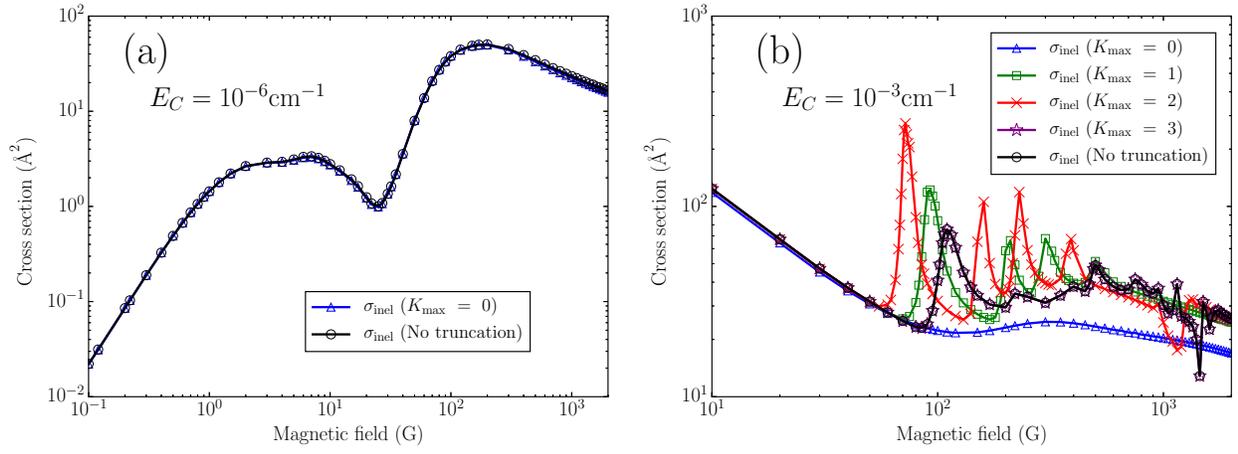}
	\renewcommand{\figurename}{Fig.}
	\caption{Magnetic field dependence of spin relaxation cross sections for Li$(\uparrow)$~+~CaH$(\uparrow)$ collisions at $E_C=10^{-6}$ cm$^{-1}$ (a) and $E_C=10^{-3}$ cm$^{-1}$ (b). Full CC results (circles) are compared with restricted basis set calculations with $K_\text{max}=0$ (triangles), 1 (squares), 2 (crosses), and 3 (stars). }
	\label{fig:licahB}
\end{figure}

Figure~\ref{fig:licahB} shows the magnetic field dependence of the inelastic cross sections for Li($\uparrow$)+CaH($\uparrow$) collisions at two representative collision energies. In the $s$-wave regime ($E_C=10^{-6}$~K), dominated by a single $l=0$ incident partial wave, the cross sections increase at low magnetic fields, reach a minimum near $B=20$~G and continue to increase at higher magnetic fields. 
The restricted basis set calculations with $K_\text{max}=0$ are in excellent agreement with full CC results over the entire range of magnetic fields.

A different trend is observed in the multiple partial wave regime ($E_C=10^{-3}$~K) in Fig.~\ref{fig:licahB}~(b). 
{While the inelastic cross section displays  a smooth magnetic field dependence below 50 G, a resonance pattern is observed at higher magnetic fields.}
The figure shows that the convergence with respect to $K_\text{max}$ slows down at higher magnetic fields, and especially in the vicinity of scattering resonances. 
Restricted basis set calculations using small basis sets do not reproduce the details of the resonance structure, which requires basis sets with $K_\text{max}\ge 3$, at which point restricted basis set calculations become nearly  as computationally expensive as full CC calculations (see Table~II).

The slow  convergence with respect to $K_\text{max}$  indicates that the matrix elements between the $K$-states omitted from the basis set are essential for the proper description of the resonance structure in Fig.~\ref{fig:licahB}(b).
 As discussed at the end of Sec. IIC, these matrix elements can involve either (1) the spin-rotation interaction (Eq.~\ref{SRint}), (2) the magnetic dipole-dipole interaction, (3) the atom-molecule interaction potential (Eq.~\ref{meV}), and (4) the squared orbital angular momentum operator (or the centrifugal term, see Eq.~\ref{meL2}).

\begin{figure}[h]
	\centering
	\includegraphics[width=1.1\textwidth, trim = 100 80 0 0]{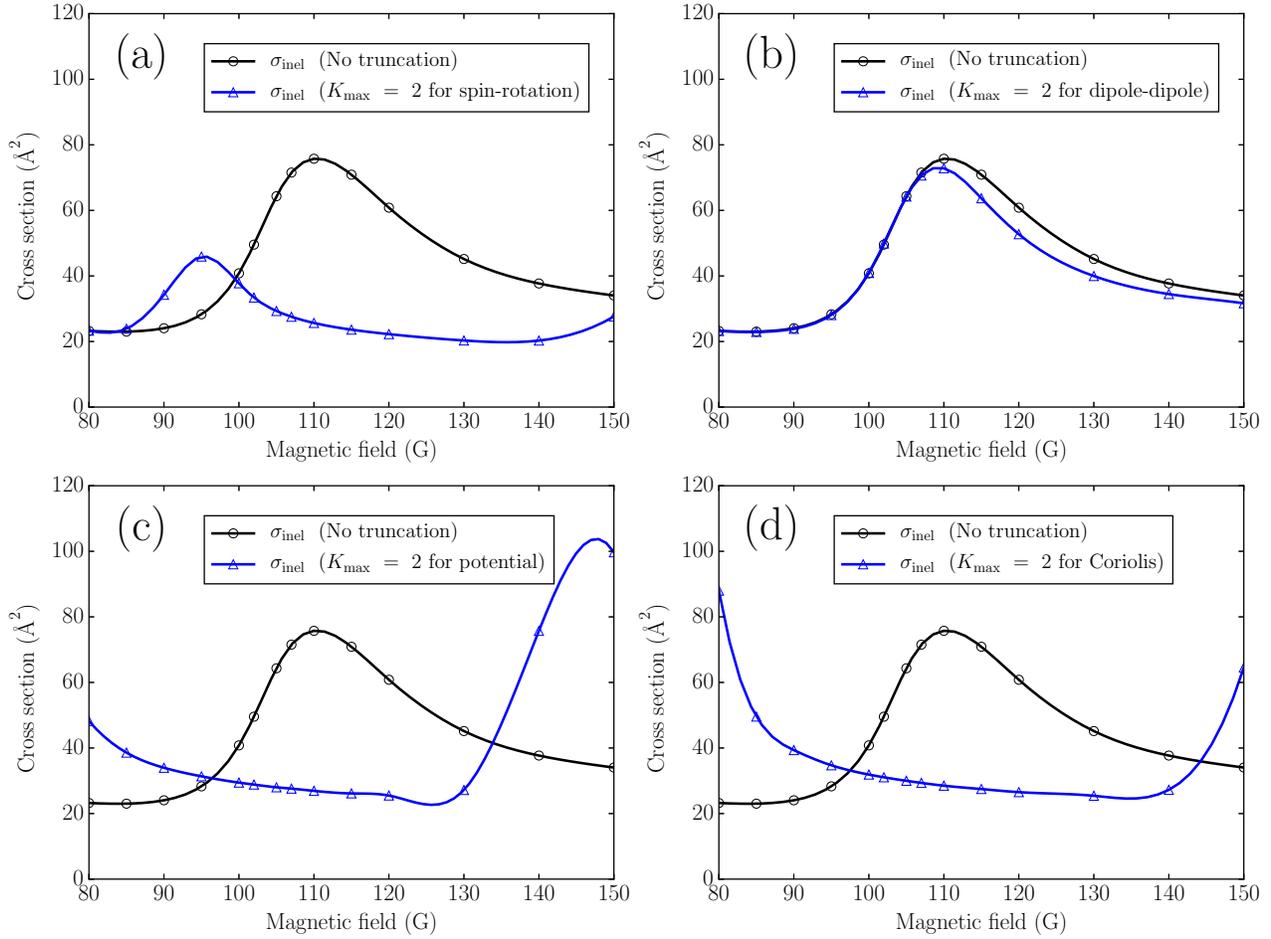}
	\renewcommand{\figurename}{Fig.}
	\caption{Magnetic field dependence of  spin relaxation cross sections for Li$(\uparrow)$~+~CaH$(\uparrow)$ collisions at $E_C=10^{-3}$ cm$^{-1}$ calculated with the $K>K_\text{max}=2$  matrix elements of the following terms omitted  from the Hamiltonian: (a) spin-rotation interaction, (b) magnetic dipole-dipole interaction, (c) atom-molecule interaction potential, and (d) centrifugal term. Full CC results (circles) are also shown for comparison.} 
	\label{fig:licahBanalysis_1}
\end{figure}

To identify which of these terms is responsible for the resonance structure shown in Fig.~\ref{fig:licahB}(b), we carried out test calculations omitting the matrix elements of the specific terms (1)-(4)  one by one. 
Figure~\ref{fig:licahBanalysis_1}(b) shows that neglecting the matrix elements of the magnetic dipole-dipole interaction between the basis states with $K=3$ has little effect on the resonance profile of the inelastic cross section. In contrast, as shown in Figs.~\ref{fig:licahBanalysis_1}(a), (c) and (d), omitting the matrix elements of the spin-rotation, interaction potential, or centrifugal terms between the basis states with $K=3$ leads to a dramatic change of the resonance profiles. This indicates that all of these interactions play an essential role in determining the resonance structure. Particularly crucial are the diagonal matrix elements of the interaction potential between  $K=3$ basis functions.



\begin{figure}[h]
	\centering
	\includegraphics[width=1.1\textwidth, trim = 100 80 0 0]{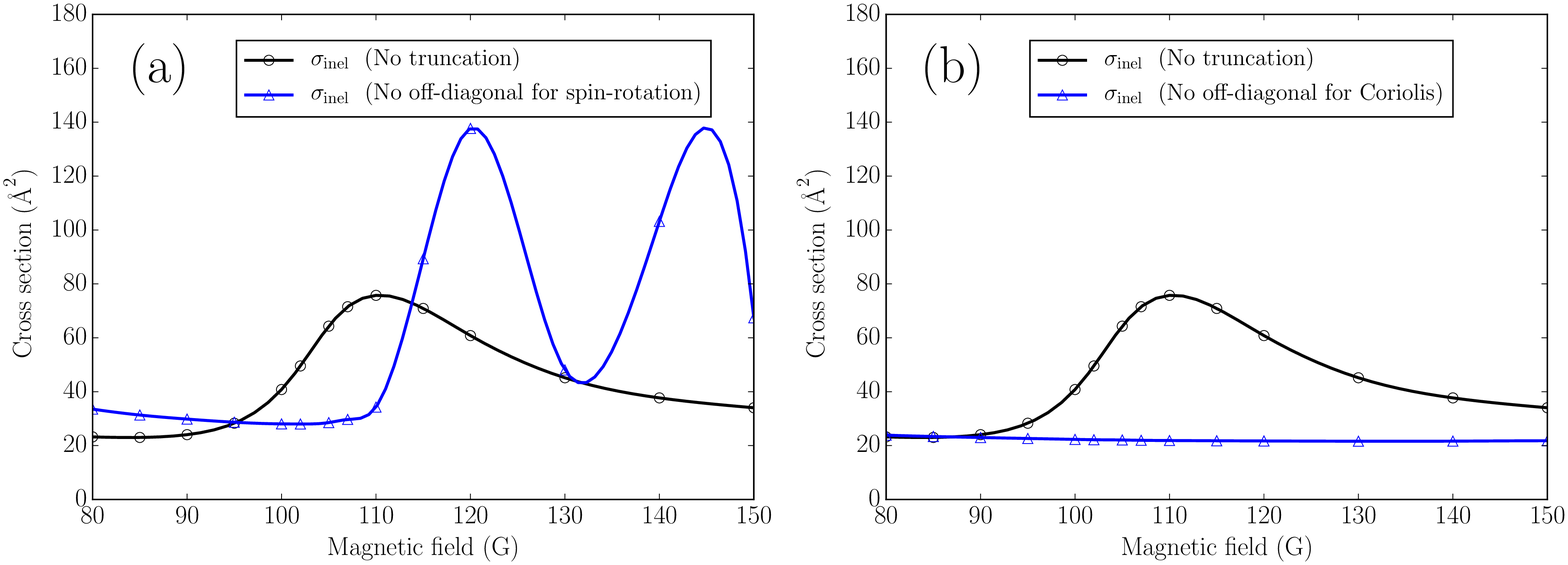}
	\renewcommand{\figurename}{Fig.}
	\caption{Magnetic field dependence of  spin relaxation cross sections for Li$(\uparrow)$~+~CaH$(\uparrow)$ collisions at $E_C=10^{-3}$ cm$^{-1}$ calculated without the off-diagonal in $K$ matrix elements of (a) the spin-rotation interaction and (b) the centrifugal term. Full CC results (circles) are also shown for comparison.  }
	\label{fig:licahBanalysis_2}
\end{figure}

The error due to the truncation of the helicity basis set may be further subdivided into the errors due to the neglect of the {\it diagonal} and {\it off-diagonal} matrix elements between the basis functions with $K>K_\text{max}$ of the spin-rotation and centrifugal interactions. 
To eliminate the diagonal error for these terms, we performed additional calculations using a complete helicity basis set but neglecting the off-diagonal couplings between the states of different $K$, thereby invoking the helicity-conserving CS approximation\cite{Kouri,Pack}.
The results  are shown in Fig.~\ref{fig:licahBanalysis_2}. We observe that the resonance profiles are sensitive to the {\it off-diagonal} matrix elements of the spin-rotation interaction and those of the centrifugal term. Remarkably, while omitting the off-diagonal matrix elements of the spin-rotation interaction splits the resonance peak in two narrower peaks, neglecting the Coriolis coupling completely sweeps out the resonance profile as seen in Fig.~\ref{fig:licahBanalysis_2}(b). This is in line with a previous CS study of the chemical reaction F+HCl, which demonstrated the inability of the CS approximation to describe the resonance structure in low-energy reaction cross sections \cite{BalaFHCl}. 
We conclude that a fully quantitative description of the magnetic resonance structure in  Li~+~CaH spin relaxation collisions  requires not only a nearly-complete helicity basis, but also the proper inclusion of all couplings between the different $K$ states in the scattering Hamiltonian.

\subsection{Li~+~SrOH collisions}

Thus far we have applied helicity basis truncation to ultracold collisions of S-state atoms with a light molecule (CaH). The spectrum of rotational levels of CaH is relatively sparse compared to that of the heavier molecular radicals slowed and trapped in recent experiments, such as SrOH \cite{Ivan} and SrF \cite{DeMille_MOT, DeMille_Btrap}. To investigate the performance of restricted helicity basis sets for collisions of heavy molecules with a dense rotational spectrum, we consider ultracold Li($\uparrow$)+SrOH($\uparrow$) collisions in a magnetic field.  The ground $^2\Sigma$ electronic state of  SrOH  has been Sisyphus  cooled \cite{Ivan} and our recent CC  study of Li~+~SrOH collisions has established  good prospects of sympathetic cooling of  SrOH$(\uparrow)$ with Li$(\uparrow)$ atoms in a magnetic trap \cite{pra17}.

\begin{figure}[h]
	\centering
	\includegraphics[width=1.1\textwidth, trim = 100 80 0 0]{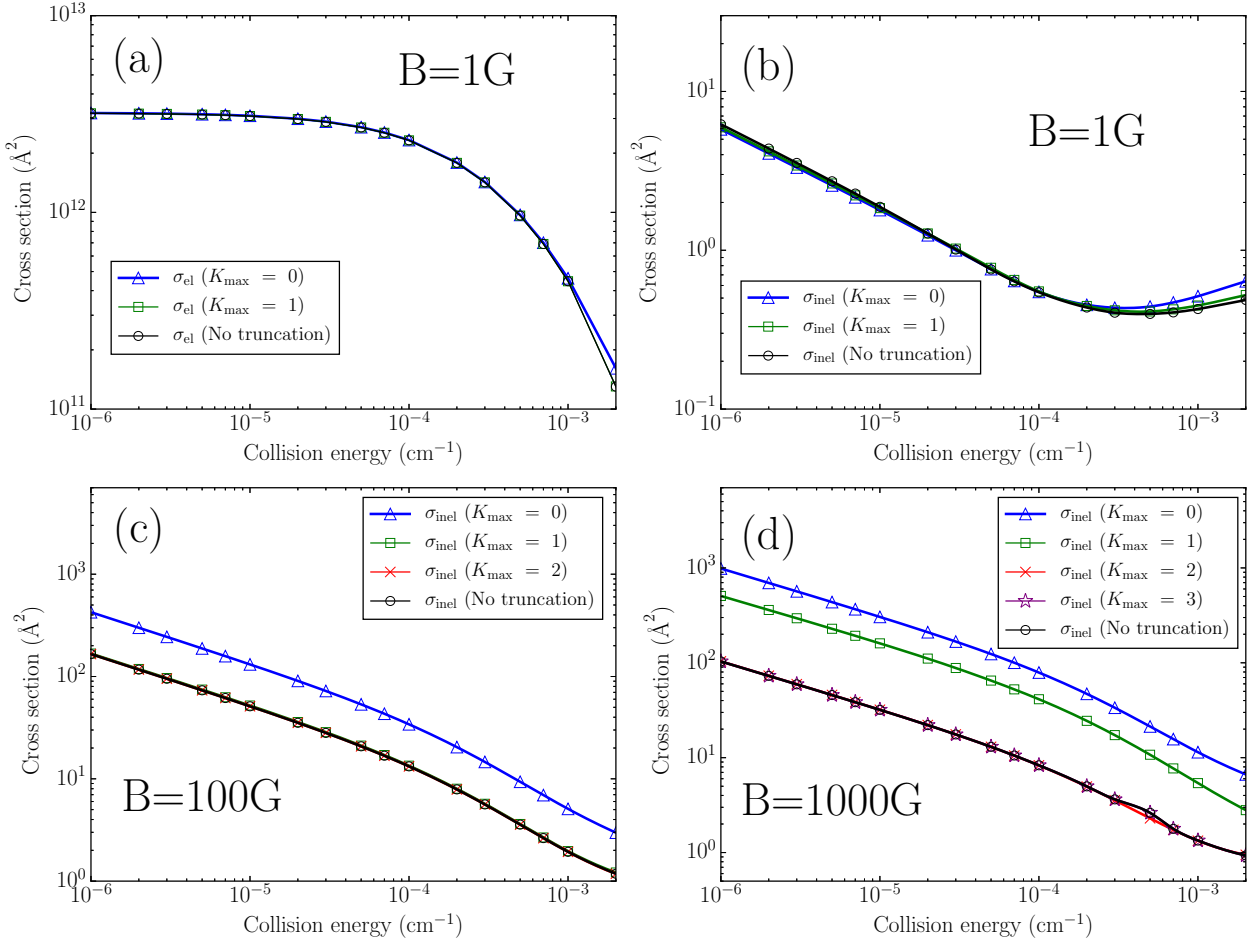}
	\renewcommand{\figurename}{Fig.}
	\caption{Elastic (a) and spin relaxation (b)-(d) cross sections for  Li$(\uparrow)$~+~SrOH$(\uparrow)$  collisions as a function of the collision energy for $B=1$~G (a)-(b), $B=100$~G (c), and $B=1000$~G (d). Full CC results (solid black line) are compared with restricted basis set calculations with $K_\text{max}=1$ (squares) and $K_\text{max}=0$ (triangles). }
	\label{fig:lisrohE}
\end{figure}

Figure~\ref{fig:lisrohE} compares the elastic and inelastic cross sections for Li$(\uparrow)$~+~SrOH$(\uparrow)$ collisions calculated using truncated basis sets with the full CC results. As in the cases of He~+~CaH  and Li$(\uparrow)$~+~CaH$(\uparrow)$ collisions considered above, the elastic cross sections obtained using the minimal basis set ($K_\text{max}=0$)  are in excellent agreement with  full CC results over the entire range of collision energies from 10$^{-3}$ cm$^{-1}$ down to the s-wave regime. At low magnetic fields, the same is true for the inelastic cross sections  shown in Fig.~\ref{fig:lisrohE}(b). However, the situation changes when the magnetic field is increased: As shown in Fig.~\ref{fig:lisrohE}(c), minimal-basis-set results  overestimate the accurate CC cross section by a factor of 2-3 depending on the collision energy. Obtaining fully converged results at higher magnetic fields ($B=1000$~G) requires larger  helicity basis sets with $K_\text{max}\ge2$. The  computational  gain achieved in this case is  modest (see Table~I).

Figure~\ref{fig:lisrohB} shows the magnetic field dependence of the cross sections for spin relaxation in ultracold Li($\uparrow$)~+~SrOH($\uparrow$) collisions in the $s$-wave regime. The cross sections increase monotonously with magnetic field at low fields ($B <500$~G)  and display a resonance structure at higher fields. The background values of the inelastic cross sections at low fields can be accurately described with $K_\text{max}\le 1$ basis sets at only a small fraction of the computational cost of the full CC calculations (see Table~I).
 Reproducing the  details of the  resonance structure  in the inset of Fig.~\ref{fig:lisrohB} requires larger basis sets $K_\text{max}\ge 2$, suggesting the importance of the rotational and Coriolis interactions in determining the resonance structure (see Sec.~IIIB above).


It is instructive to compare the overall performance of restricted helicity basis sets applied to Li~+~SrOH vs Li~+~CaH collisions. Despite the very disparate rotational constants of CaH ($B=4.2$ cm$^{-1}$) and SrOH ($B=0.25$ cm$^{-1}$) the Li-SrOH and Li-CaH collision complexes have similar reduced masses and hence similar  Coriolis interactions, leading one to expect no significant differences in performance.  Comparing Fig.~\ref{fig:licahE} and Fig.~\ref{fig:lisrohE} we see that this is indeed the case: In both cases relatively  large basis sets with $K_\text{max}\ge 2$ are required to properly describe the resonance structure at high magnetic fields. This observation suggests that restricted basis set  calculations might provide more accurate results for heavier atom-molecule collisions (such as Rb~+~SrF \cite{pra18subm}) where the Coriolis interaction is suppressed by the large reduced mass of the collision complex.

\section{Summary and outlook}

We have explored the use of truncated helicity basis sets to enhance the computational efficiency of quantum scattering calculations of ultracold atom-molecule collisions in the presence of an external magnetic field. We have demonstrated that accurate elastic and inelastic cross sections for cold He~+~CaH, Li~+~CaH, and Li~+~SrOH collisions can be obtained with  a limited number of $K$-states in the basis set. This leads to substantial, 10-100 fold gains in computational efficiency over the full CC calculations (see Table~II) for heavy, strongly anisotropic atom-molecule collision complexes such as Li-CaH and Li-SrOH.

\begin{figure}[h]
	\centering
	\includegraphics[width=0.7\textwidth, trim = 0 0 0 0]{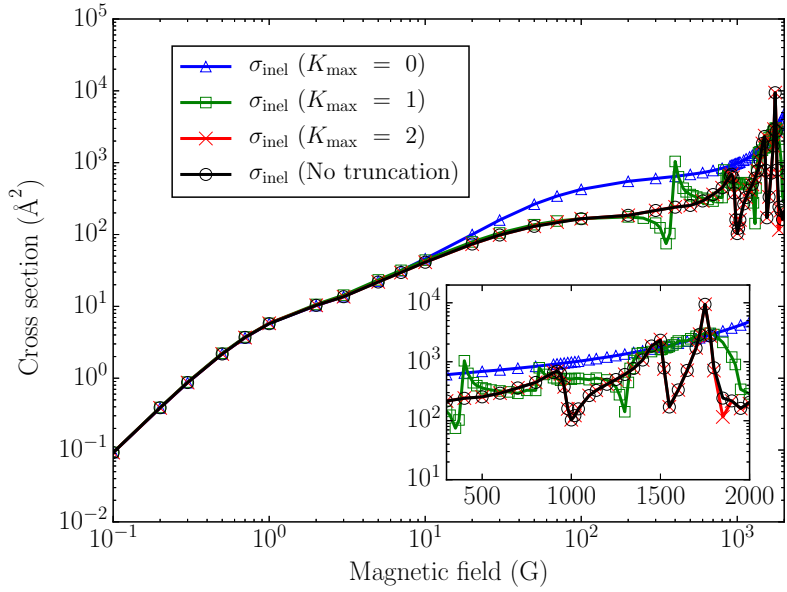}
	\renewcommand{\figurename}{Fig.}
	\caption{Magnetic field dependence of spin relaxation cross sections for Li$(\uparrow)$~+~SrOH$(\uparrow)$ collisions at $E_C=10^{-6}$ cm$^{-1}$. Full CC results (solid line) are compared with restricted basis set calculations (symbols) with $K_\text{max}=0$ (triangles), 1 (squares), 2 (crosses), and 3 (stars). The inset shows the details of the resonance structure at higher magnetic fields.}
	\label{fig:lisrohB}
\end{figure}

The rate of convergence of restricted basis set calculations towards the exact (full CC) results is an important factor that determines the computational advantage of using truncated helicity basis sets. We found that this rate  generally depends on the collision system under consideration, the collision energy and on the magnitude of the external magnetic field. The fastest convergence is observed for elastic cross sections: As shown in Figs.~\ref{fig:hecahE}, \ref{fig:licahE}, and \ref{fig:lisrohE}, quantitative accuracy can be achieved away from the resonance region with a minimal basis set ($K_\text{max}=0$) over wide range of collision energies for all of the collision systems studied here. The spin relaxation cross sections for Li~+~CaH and Li~+~SrOH in the $s$-wave regime also display extremely rapid convergence with respect of $K_\text{max}$  as shown in Figs.~\ref{fig:licahB}(a) and \ref{fig:lisrohB}. In these regime, we achieve computational gains on the order of 10-100 (see Table~II).  The reason for such rapid convergence is that the main mechanisms responsible for either elastic scattering or dipolar-medicated spin relaxation in  Li~+~CaH and Li~+~SrOH collisions are diagonal in $K$. Larger  basis sets are required to describe collisional relaxation mechanisms that do not conserve $K$, such as the intramolecular spin-rotation interaction, which is responsible for spin relaxation in He~+~CaH collisions [Fig.~\ref{fig:hecahE}(b)].

Another instance where the $K$ non-conserving mechanisms are important is provided by the magnetic resonance structure in Li~+~CaH and Li~+~SrOH collisions in the few-partial wave regime  [Figs.~\ref{fig:licahB}(b) and \ref{fig:lisrohB}]. These resonances arise as a result of an interplay between the atom-molecule interaction potential, the spin-rotation interaction, and the  centrifugal interaction,
which have both the diagonal and off-diagonal matrix elements in $K$ (except for the interaction potential). Excluding the basis functions with $K\ge K_\text{max}$  from the basis eliminates the essential matrix elements from the matrix representation of the Hamiltonian, leading to either a substantial modification (for small $K_\text{max})$ or complete disappearance (for $K_\text{max}=0$) of the resonance structure as shown in  Figs.~\ref{fig:licahB}(b) and \ref{fig:lisrohB}. We conclude that the accurate description of the resonance structure as a function of either collision energy or magnetic field requires nearly complete helicity basis sets with $K_\text{max}$ approaching the maximal possible values of $J$ included in the basis set.

However, if the accurate description of the resonance positions and widths is not required, the background values of the inelastic cross sections can be estimated in calculations employing minimal basis sets [see Figs.~\ref{fig:licahB}(b) and \ref{fig:lisrohB}.]

In summary, our results demonstrate that CC calculations employing restricted helicity basis sets can be used to obtain accurate atom-molecule scattering cross sections over a wide range of collision energies and magnetic fields at a fraction of the computational cost of the  fully rigorous CC calculations (see Table~II).  However, this technique converges much more slowly in the presence of scattering resonances, and hence must be used with caution when applied to calculate the resonance positions and widths.

In future work, it would be interesting to  explore the performance of truncated helicity basis sets in CC calculations of heavy atom-molecule collisions  (such as Rb-SrF \cite{pra18subm}). The weaker  Coriolis interactions in these systems may lead to better performance of truncated basis sets. This  methodology may also be applied to the computationally challenging problems of atom-molecule reactive scattering   or molecule-molecule collisions in external electromagnetic fields \cite{jcp12,prl15}.


\section*{Acknowledgements}
We are grateful to Naduvalath Balakrishnan for a helpful discussion and to Roman Krems for providing access to  additional computational resources. This work was supported by NSF grant No. PHY-1607610.

\newpage

\begin{table}[h]
\caption{Convergence parameters of CC calculations (values of $R$ are given in $a_0$). The reduced masses of the He-CaH, Li-CaH, and Li-SrOH collision complexes used in CC calculations were 2.8138538, 5.9902077 and 6.5762047 a.m.u. }
\begin{tabular} {cccccc}  
\hline \hline
System        & \  $R_\text{min}$ \  & \ $R_\text{max}$ \ & \ $\Delta R$ \ & \ $N_\text{max}$ \  &  \ $J_\text{max}$ \   \\  
\hline\hline
\ He~+~CaH   \          &  \  2.0  \   &  \ 100.0  \ & \ 0.04  \ &  4  &   4.5    \\
\ Li~+~CaH     \          &  \  3.78 \  &  \ 944.9 \   & \  0.0038  \ &  55  &   3    \\
\ Li~+~SrOH   \          &   \ 7.56  \ &  \ 1133.8  \ &  \ 0.0019  \ & 115  &   3            \\
\hline
\end{tabular}
\label{tab:parameters}
\end{table}

\newpage

\begin{table}[h]
\caption{Numbers of scattering channels $N_c$ for the full and truncated basis sets. The last column lists the computational gain achieved by using the truncated basis set.    }
\begin{tabular} {ccccccc}  
\hline \hline
System       &   $J_\text{max}$        &    $N_\text{max}$     &  \ \ $M$ \ \ &     $K_\text{max}$    &  $N_c$   &    ({$N_c^\text{full}$}/$N_c$)$^3$   \\  
\hline\hline
{He~+~CaH}  &  {4.5}            &    {4} 			&    1/2 			& 4                           &  {190}  &   1     \\
                                      &                                        &     	                              &                &  3                          &  {184} & {1.10}     \\
                                      &                                        &     	                              &                &  2                          &  {164} &  {1.55}     \\
                                      &                                        &     	                              &                &  1                          &  {122} &  {3.78}     \\
                                      &                                        &     	                              &                &  0                          &  {50}   &  {54.87}     \\
\hline
Li~+~CaH   &          3            &   55   &   0     &   4   &  3496 &  1     \\
                   &                        &          &          &   3   &  3392 &  1.09     \\
                   &                        &          &          &   2   &  2968 &  1.63     \\
                   &                       &          &          &   1   &  2104 &  4.59     \\
                   &                       &          &          &   0   &   784 &  88.67     \\
\hline
Li~+~SrOH   &          3            &   115   &  0   &   4   &  7336 &  1     \\
                  &              &          &                      &   3   &  7112 &  1.10    \\
                  &              &          &                      &   2   &  6208 &  1.65     \\
                  &              &          &                      &   1   &  4384 &  4.69     \\
                  &              &          &                     &   0   &   1624 &  92.18     \\
\hline
\end{tabular}
\label{tab:parameters}
\end{table}

\newpage

\newpage

\newpage

\newpage

\newpage

\newpage

\newpage

\end{document}